\documentclass[cameraready]{Interspeech}

\title{TriA Pipeline: A Large-Scale Automatic Audio Annotation Pipeline For Audio Classification In Specific Scenarios}

\author[orcid=0009-0000-1944-2798]{Hong}{Lyu}
\author[orcid=0009-0003-1656-7773]{Mingru}{Yang}
\author[orcid=0000-0002-9079-4566, correspondingauthor]{Qianhua}{He}
\author[orcid=0000-0003-4362-1125, correspondingauthor]{Yanxiong}{Li}
\author[]{Jinxin}{Huang}
\author[]{Zhengyu}{Pei}

\address{
    School of Electronic and Information Engineering, South China University of Technology, Guangzhou, China
}

\email{eehlyu@mail.scut.edu.cn, eeqhhe@scut.edu.cn, eeyxli@scut.edu.cn}


\usepackage{comment}
\usepackage{makecell}
\usepackage{multirow}

\begin{document}

\maketitle

\begin{abstract}
    There are some datasets of varying scales for audio classification (AC) applied to different tasks. However, annotated data is limited for most scenarios, such as domestic environments. To address this challenge, we propose an \textbf{A}utomatic \textbf{A}udio \textbf{A}nnotation Pipeline--TriA Pipeline, which can efficiently convert audio from various scenarios into high-quality training data with audio event annotations. A TriA dataset was constructed with the TriA Pipeline, over 2130 hours of audio covering 431 audio classes. Furthermore, we partitioned a prior-knowledge-guided subset (TriA$_{\mathrm{GK}}$) from TriA and conduct comparative experiments on three domestic AC tasks. Comparing the result on manually annotated data only and that on manually annotated data combines TriA$_{\mathrm{GK}}$, TriA$_{\mathrm{GK}}$ could achieve average relative gains of 3.97\% in accuracy and 3.35\% in Macro-F1, validating the effectiveness of TriA$_{\mathrm{GK}}$ and the TriA Pipeline.

\end{abstract}

\section{Introduction}

Audio Classification (AC) enables the recognition of various environmental sound events, serving as a core component in applications ranging from multimedia content analysis \cite{gemmeke2017audio} and audio captioning \cite{drossos2020clotho} to bio-acoustic monitoring \cite{kvsn2020bioacoustics, terenzi2021comparison}. Ultimately, the efficacy of these AC systems heavily depends on the availability of large-scale audio datasets that cover diverse acoustic scenes.

Existing AC datasets can be broadly categorized into general-purpose (e.g., AudioSet \cite{gemmeke2017audio}, FSD50K \cite{fonseca2021fsd50k}, ESC-50 \cite{piczak2015dataset}, FSC-89 \cite{11164310}) and specialized ones (e.g., Kitchen20 \cite{moreaux2019benchmark}, CHiMe-Home \cite{foster2015chime}, NSynth-100 \cite{11457822}). The former encompass a wide variety of acoustic scenes, while the latter focus on specific domains, such as domestic environments or human non-speech sounds. Among general-purpose AC datasets, AudioSet is the most extensive,  consisting of a class-balanced subset (AS-20K), a class-unbalanced subset (AS-2M), and an evaluation set, with AS-2M being widely utilized for pre-training and fine-tuning audio models \cite{kong2020panns, chen2023beats, li2022atst, dinkel2024streaming}. FSD50K is another large-scale audio dataset, but remains class-unbalanced, leaving certain sound classes underrepresented, such as wails, moans, wheezes, squeals, and specific domestic sounds. ESC-50, by contrast, is a widely used class-balanced dataset of 50 classes across 5 major categories, though its scale is notably limited. In summary, existing general-purpose AC datasets suffer from two primary limitations: (i) insufficient data for specific acoustic scenes, and (ii) limited dataset scale. 

Among specialized AC datasets, DESED \cite{turpault:hal-02160855} comprises real recordings (DESED${\mathrm{real}}$) and synthetic data, covering 10 classes of domestic audio events, and is widely used for AC and sound event detection (SED) in domestic scenes \cite{9414643, khandelwal23_interspeech, khandelwal2023leveraging}. However, DESED$_{\mathrm{real}}$ contains only 5955 annotated clips. Kitchen20 is designed for kitchen AC tasks, while both HTAD \cite{garcia2021htad} and CHiMe-Home target domestic activity recognition. CIRDO \cite{vacher2016cirdo} and BiMP \cite{dibble2025bi} are simulated datasets for safety monitoring of elderly individuals living alone, and Nonspeech7k \cite{rashid2023nonspeech7k}, originally developed for paralinguistic classification, can similarly be applied to domestic safety monitoring. Nevertheless, all these specialized datasets suffer from limited scale. In general, annotated audio data remains scarce across many scenarios, with the problem becoming more pronounced in highly specific domains.

To address the scarcity of annotated audio data in specific scenarios, we propose a large-scale \textbf{A}utomatic \textbf{A}udio \textbf{A}nnotation Pipeline--TriA Pipeline. TriA Pipeline consists of four stages: Standardization, Audio Activity Detection (AAD), Audio Event Detection (AED), and Filtering, which collectively convert raw audio from diverse streaming platforms and scenarios into high-quality training data with audio event annotations. The most closely related works are Emilia-Pipe \cite{he2024emilia}, NVSpeech-Pipe \cite{liao2025nvspeech}, and NonVerbalSpeech-Pipe \cite{ye2025scalable}. However, Emilia-Pipe relies on Automatic Speech Recognition (ASR) and supports only speech data annotation, while both NVSpeech-Pipe and NonVerbalSpeech-Pipe focus exclusively on paralinguistic speech annotation. In contrast, by integrating an AED module, TriA Pipeline supports audio event annotation across a broad range of scenarios, resolving the annotation scarcity problem in the specific domains described above. Both objective metrics and subjective listening evaluations confirm that TriA Pipeline produces high-quality and diverse audio data with high annotation reliability.

Based on TriA Pipeline, the TriA dataset is further constructed, containing over 2130 hours of high-quality audio data covering 431 audio classes. To evaluate the effectiveness of TriA, we partition a prior-knowledge-guided subset, TriA$_{\mathrm{GK}}$, from TriA and setup three specific classification tasks: DESED for Audio Classification (DESED$_{\mathrm{AC}}$), Kitchen20, and Nonspeech7k. They represent three domestic AC tasks: general domestic AC, kitchen AC, and domestic safety monitoring. We conduct comparative experiments using TriA$_{\mathrm{GK}}$ and manually annotated datasets. Experimental results show that fine-tuning models using only TriA$_{\mathrm{GK}}$ achieves performance comparable to models fine-tuned using manually labeled data. Furthermore, fine-tuning the model by combining TriA$_{\mathrm{GK}}$ with manually annotated data can achieve average relative improvements of 3.97\% in accuracy and 3.35\% in Macro-F1. These results indicate that TriA$_{\mathrm{GK}}$ can help the model achieve better performance in the specified classification task, and further validate the effectiveness of the proposed TriA Pipeline.

\begin{figure*}[t]
    \centering
    \includegraphics[width=0.98\linewidth]{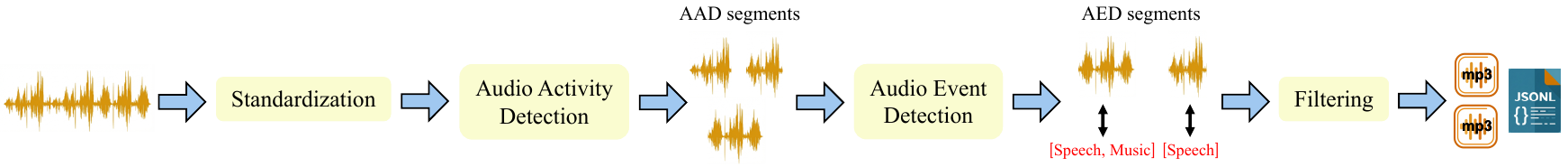}
    \caption{Overview of the TriA Pipeline. AAD and AED denote Audio Activity Detection and Audio Event Detection.}
    \label{fig:1}
\end{figure*}

\section{TriA Pipeline}

This section details the TriA Pipeline and its evaluation on minibatch data. As illustrated in Figure~\ref{fig:1}, the TriA Pipeline consists of four main stages: Standardization, Audio Activity Detection (AAD), Audio Event Detection (AED), and Filtering.

\subsection{Standardization}

This step follows the same procedure as Emilia-Pipe and aims to standardize audio with heterogeneous formats for subsequent processing. Specifically, the original audio recordings are converted into mono-channel WAV format with a sample rate of 24 kHz and a 16-bit sample width. The target loudness is then normalized to -20 dBFS, while the signal amplitude is constrained within the range of -3 dB to 3 dB to prevent distortion. Finally, each waveform is normalized by dividing all sample points by the maximum amplitude.

\subsection{Audio Activity Detection}

The purpose of this step is to split long audio recordings into clips of appropriate length while removing redundant segments. The auditok tool\footnote{\href{https://github.com/amsehili/auditok}{\url{https://github.com/amsehili/auditok}}} is used to split each sample into AAD segments based on a preset energy threshold, with the maximum segment duration constrained to 30 s. 

To determine the appropriate minimum duration for sound activity segments and maximum duration for silent segments, subjective listening tests can be conducted on datasets related to specific scenarios. For each audio class in each dataset, 10 instances are randomly selected for listening. During the tests, the minimum duration required to identify the event in each instance and the temporal interval between two adjacent events of the same class are recorded. Based on the test results, the average minimum time required to identify events of each class is calculated and referred to as the Event Critical Time (ECT). The average interval between adjacent events of each class is also calculated and referred to as the Silent Critical Time (SCT). During the AAD processing, retained activity segments are required to exceed the minimum ECT, as shorter segments contain insufficient information to identify. The retained silent segments are restricted to be shorter than the maximum SCT, since excessively long silent segments introduce redundant information and reduce efficiency. In our experiments, the target scenario is the domestic scene, where the minimum ECT and the maximum SCT are 1.2 s and 2.0 s, respectively.

\subsection{Audio Event Detection}

To enable the TriA dataset to be directly applied to AC tasks, the BEATs model\footnote{\url{https://github.com/microsoft/unilm/tree/master/beats}} is employed to annotate audio events for AAD segments. The BEATs model achieves state-of-the-art (SOTA) performance on the AS-2M dataset under the single audio modality and can detect 527 audio classes \cite{chen2023beats}.

Specifically, for each AAD segment, the AS-2M fine-tuned BEATs$_{\mathrm{iter3+}}$ model is used for event detection, dividing it into multiple segments with event annotations. To detect short events in the segment, local detection is conducted by scanning each AAD segment with a fixed detection window length and window shift, producing preliminary detection segments. Adjacent detection segments within the same AAD segment are then concatenated if their annotated Top-1 events are identical. To further detect long audio in the segment and continuous event across adjacent detection windows, global detection is performed on each concatenated detection segment. In global detection, the window length is equal to the length of the segment to be detected. If the globally detected class differs from the original class, the newly class is appended to the segment annotation. The resulting segments are called AED segments.

The window length for AED local detection is required to exceed the maximum ECT (3 s). The purpose is to enable the model to process one or more target events as completely as possible in a single detection, providing sufficient information to the model and improving the reliability of the detection results. However, if the window length is too long, the audio segment detected by the model at one time may include too many events, which will bring difficulties to the model detection. The higher the event confidence threshold for AED detection, the higher the reliability of the detection results. In our experiment, the event confidence threshold is set to 0.6. The window length and window shift for local detection are set to 5 s and 3 s, respectively.

\begin{table*}[t]
    \centering
    \caption{Statistical results of 285 hours of original data processed by the TriA Pipeline. Filtering$_\mathrm{1}$ uses PC, PQ, and CLAP similarity thresholds of 1.8, 5.5, and 5, respectively, while Filtering$_\mathrm{2}$ uses thresholds of 2.24, 5.85, and 7.43, respectively.}
    \label{tab:1}
    \vspace{-0.2cm}
    \resizebox{\textwidth}{!}{%
    \begin{tabular}{@{}cccccccccc@{}}
        \toprule
        \textbf{Data} & \multicolumn{3}{c}{\textbf{Duration (s)}} & \multicolumn{2}{c}{\textbf{Aesthetics PC $|$ PQ}} & \multicolumn{2}{c}{\textbf{CLAP Similarity}} & \multirow{2}{*}{\textbf{\makecell{Total\\ Duration (hours)}}} & \multirow{2}{*}{\textbf{\makecell{Total\\ Classes}}} \\ 
        \cmidrule(lr){2-4} \cmidrule(lr){5-6} \cmidrule(lr){7-8}
        & \textbf{min} & \textbf{max} & \textbf{avg} \(\pm\) \textbf{std} & \textbf{min} & \textbf{avg} \(\pm\) \textbf{std} & \textbf{min} & \textbf{avg} \(\pm\) \textbf{std} & & \\ 
        \midrule
        Original Audio & 2.04 & 17605.37 & 233.05 \(\pm\) 841.70 & 1.39 $|$ 3.12 & 5.17 \(\pm\) 1.66 $|$ 7.20 \(\pm\) 0.89 & — & — & 284.70 (100.00\%) & — \\
        Annotated w/o Filtering & 1.20 & 30.00 & 10.87 \(\pm\) 9.22 & 1.33 $|$ 2.85 & 4.60 \(\pm\) 1.71 $|$ 6.52 \(\pm\) 1.07 & -4.19 & 7.71 \(\pm\) 2.52 & 242.62 (85.22\%) & — \\
        Annotated w Filtering$_\mathrm{1}$ & 1.20 & 30.00 & 11.57 \(\pm\) 9.38 & 1.80 $|$ 5.50 & 4.90 \(\pm\) 1.62 $|$ 6.92 \(\pm\) 0.77 & 2.00 & 8.04 \(\pm\) 2.24 & 182.37 (64.06\%) & 325 \\
        Annotated w Filtering$_\mathrm{2}$ & 1.20 & 30.00 & 11.70 \(\pm\) 9.83 & 2.24 $|$ 5.85 & 5.18 \(\pm\) 1.45 $|$ 7.10 \(\pm\) 0.71 & 7.43 & 9.54 \(\pm\) 1.65 & 80.08 (28.13\%) & 258 \\ \bottomrule
    \end{tabular}%
    }
\end{table*}

\subsection{Filtering}

The original audio may exhibit varying quality and the BEATs model may produce detection errors. To ensure the data quality and improve the matching degree between data and event annotations, the audiobox-aesthetics\footnote{\href{https://github.com/facebookresearch/audiobox-aesthetics}{\url{https://github.com/facebookresearch/audiobox-aesthetics}}} and the CLAP model\footnote{\href{https://github.com/microsoft/CLAP}{\url{https://github.com/microsoft/CLAP}}} are used to filter AED segments. Specifically, Production Complexity (PC) and Production Quality (PQ) in aesthetics are used as filtering indicators. PC and PQ are relatively objective indicators, focusing on the complexity of the audio scene and the technical quality respectively \cite{tjandra2025aes}. For each AED segment, the PC and PQ, as well as the CLAP similarity between the segment and event annotation \cite{CLAP2022}, are calculated. Segments with PC, PQ, or CLAP similarity below predefined thresholds are filtered. The remaining high-quality segments are stored in MP3 format, and a JSONL (JSON Lines) file containing the associated metadata is generated for efficient indexing and retrieval.

\subsection{Evaluation on minibatch data}

To validate the effectiveness of each individual module and the overall TriA Pipeline, 284.7 hours of original audio are randomly collected from streaming platforms and processed using the pipeline. The experiment is conducted on an NVIDIA RTX 3090 GPU, with a total processing duration of about 10 hours and the RTF of 0.03. As shown in Table~\ref{tab:1}, statistics are performed across multiple objective evaluation perspectives. The original audio exhibits a broad duration distribution, with a relatively long average duration and a large variance. In contrast, the filtered data shows a more standardized and appropriate duration. Compared with the unfiltered data, the filtered data achieves much higher average PC, PQ, and CLAP Similarity, indicating that the data quality and annotation reliability have been significantly improved. However, the quality of the unfiltered data is lower than that of the original data. The reason is that the original data has longer average duration and higher average sampling rate, which favor PC and PQ measures. After increasing the filtering threshold, the quality and annotation reliability are further improved. However, it reduces the class diversity and the total duration of the data. In addition, subjective listening tests are also conducted, randomly selecting 100 samples to assess annotation accuracy, and an average accuracy of 93.67\% is obtained. Overall, the results demonstrate that the TriA Pipeline is feasible and effective. It can increase the amount of data in each class and continuously expand the scale of training data through repeated large-scale audio data collection and pipeline processing.

\section{TriA Dataset}

\subsection{Statistics and Analysis}

\begin{figure}[ht]
    \centering
    \includegraphics[width=0.65\linewidth]{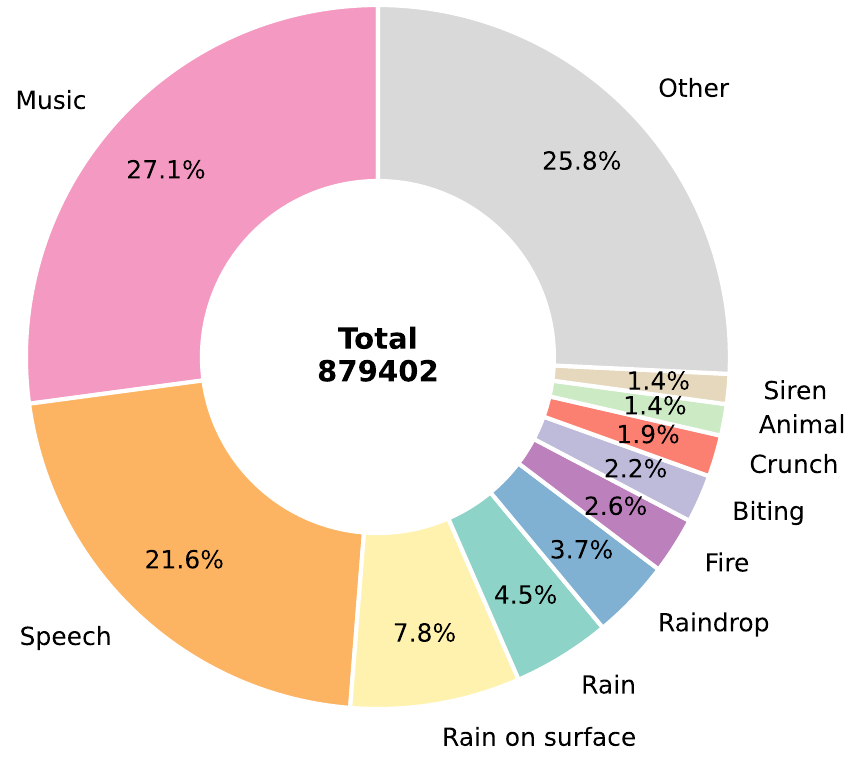}
    \caption{Class distribution of TriA.}
    \label{fig:2}
\end{figure}

Over 8706 hours of original audio data were collected from streaming platforms including Bilibili and Douyin, covering diverse topics such as daily life, entertainment media, and technology. After processing with the TriA Pipeline, the TriA dataset was constructed. The TriA dataset contains over 2130 hours of audio data, covering 431 audio classes. As illustrated in Figure~\ref{fig:2}, the class distribution of TriA is presented. Among them, the class with the largest number of samples is Music, which is attributed to the widespread presence of background music in streaming videos. Although the class distribution of the data is imbalanced, task-specific balanced subsets can be constructed for downstream applications.

To analyze the data quality and annotation reliability of the TriA dataset, Table~\ref{tab:2} summarizes statistics on PC, PQ, and CLAP similarity of different datasets. TriA achieves the best PC and PQ, indicating that the data quality of TriA is high, which benefits from the filtering stage in the TriA Pipeline. The CLAP similarity of TriA ranks third among all datasets, showing that its annotation reliability is not as good as that of Kitchen20 and Nonspeech7k, but better than that of DESED$_{\mathrm{real}}$. This difference can be attributed to the annotation method of the datasets. Kitchen20 and Nonspeech7k are manually annotated. TriA is automatically annotated using the pipeline, whereas DESED$_{\mathrm{real}}$ is annotated through crowdsourcing.

\begin{table}[t]
    \centering
    \caption{PC, PQ, and CLAP similarity of different datasets. The metrics for TriA are calculated on a randomly sampled 300-hour subset.}
    \label{tab:2}
    \vspace{-0.2cm}
    \resizebox{0.95\columnwidth}{!}{%
    \begin{tabular}{ccc}
        \toprule
        \textbf{Data} & \textbf{Aesthetics PC $|$ PQ} & \textbf{CLAP Similarity} \\ 
        \midrule
        DESED$_{\mathrm{real}}$ & 3.39 \(\pm\) 0.86 $|$ 5.85 \(\pm\) 0.85 & 7.43 \(\pm\) 4.89 \\
        Kitchen20 & 2.49 \(\pm\) 0.37 $|$ 6.38 \(\pm\) 0.74 & 12.87 \(\pm\) 3.41 \\
        Nonspeech7k & 2.24 \(\pm\) 0.67 $|$ 6.34 \(\pm\) 0.80 & 11.71 \(\pm\) 3.14 \\
        \midrule
        TriA & 5.13 \(\pm\) 1.48 $|$ 7.08 \(\pm\) 0.67 & 9.62 \(\pm\) 1.54 \\ 
        \bottomrule
    \end{tabular}%
    }
\end{table}

\subsection{Prior-knowledge-guided Subset}

The prior-knowledge refers to the threshold used for specific scenario, which was determined through subjective listening tests and statistics related to the specific scenario. The samples related to these prior-guided thresholds are collected to a subset, TriA$_{\mathrm{GK}}$.

For convenience of subsequent experiment, TriA$_{\mathrm{GK}}$ is further partitioned into three subsets: TriA$_{\mathrm{DESED}}$, TriA$_{\mathrm{Kitchen20}}$, and TriA$_{\mathrm{Nonspeech7k}}$. The audio classes they cover are the same as the classes of DESED$_{\mathrm{real}}$, Kitchen20, and Nonspeech7k respectively. The class nomenclature of the TriA follows the AudioSet ontology \cite{gemmeke2017audio}. For example, the class \textit{Electric\_shaver\_toothbrush} in DESED$_{\mathrm{real}}$ corresponds to \textit{Electric shaver, electric razor} and \textit{Electric toothbrush} in TriA. Table~\ref{tab:3} reports the number of clips and total duration for each dataset. The scale of each subset of TriA$_{\mathrm{GK}}$ is larger than that of the corresponding dataset.

\begin{table}[t]
    \centering
    \caption{Number of clips and total duration of different datasets. Unlabeled data are excluded from the statistics.}
    \label{tab:3}
    \vspace{-0.2cm}
    \resizebox{0.95\columnwidth}{!}{%
    \begin{tabular}{cccc}
        \toprule
        \multicolumn{2}{c}{\textbf{Dataset}} & \textbf{Clips} & \textbf{Total Durations (hours)} \\ 
        \midrule
        \multirow{3}{*}{\makecell{Manual\\ annotate}} & DESED$_{\mathrm{real}}$ & 5955 & 16.50 \\
        & Kitchen20 & 1070 & 1.47 \\
        & Nonspeech7k & 7014 & 6.75 \\
        \midrule
        \multirow{3}{*}{TriA$_{\mathrm{GK}}$} & TriA$_{\mathrm{DESED}}$ & 23517 & 42.60 \\
        & TriA$_{\mathrm{Kitchen20}}$ & 1688 & 2.45 \\
        & TriA$_{\mathrm{Nonspeech7k}}$ & 9618 & 11.29 \\
        \bottomrule
    \end{tabular}%
    }
\end{table}

\section{Experiments}

This section evaluates the effectiveness of TriA$_{\mathrm{GK}}$ for three domestic AC tasks: DESED$_{\mathrm{AC}}$, Kitchen20, and Nonspeech7k. For each task, three experiments are conducted. The baseline model is trained with different datasets, the first on the manually annotated data, the second on the TriA$_{\mathrm{GK}}$, and the third on both, the TriA$_{\mathrm{GK}}$ first and then the manually annotated data (refer to Table~\ref{tab:4}), to analyze whether the pipeline-processed data can help the model learn the specified AC task.

\subsection{Implementation Details}

\subsubsection{Baseline}

The BEATs model \cite{chen2023beats} is adopted. The backbone consists of a 12-layer Transformer encoder with 90M parameters, initialized with the pre-trained weights BEATs$_{\mathrm{iter3+}}$. A linear classification head, comprising two fully connected layers, is appended to the backbone. Input audio is resampled to 16 kHz. We extract 128-dimensional Mel-filter bank features using a Povey window of 25 ms and a hop size of 10 ms.

All experiments are conducted on an RTX 3090 GPU. The cross entropy loss and the AdamW optimizer are used. The maximum training epoch is set to 50, with an early stopping patience of 15. For the DESED$_{\mathrm{AC}}$ and Nonspeech7k tasks, the entire BEATs model is fully fine-tuned. For the Kitchen20 task, only the classification head is fine-tuned. The learning rates for full fine-tuning and adapter fine-tuning are set to 5e-5 and 6e-3, respectively, following a cosine annealing learning rate schedule.

Accuracy and Macro-F1 are used as evaluation metrics. Accuracy measures the overall classification correctness, while Macro-F1 reflects the consistency of model performance across different classes. The validation metric is accuracy + Macro-F1.

\subsubsection{Task Setup}

The DESED$_{\mathrm{real}}$ training set contains 4429 clips, consisting of the real strongly labeled set and the real weakly labeled set. The validation set and test set contain 373 and 1153 clips respectively, which are derived from the real strongly labeled validation set and the real strongly labeled test set. To adapt DESED$_{\mathrm{real}}$ to the AC setting, the strongly labeled sets are converted into weak labels by retaining only the audio paths and event annotations. TriA$_{\mathrm{DESED}}$ is split into training and validation sets with a 9:1 ratio. The test set for the DESED$_{\mathrm{AC}}$ task is the DESED$_{\mathrm{real}}$ test set.

Previous research evaluates Kitchen20 using 5-fold cross-validation. To unify the test set for the Kitchen20 task, the fifth fold of Kitchen20 dataset is used as the test set, which contains 160 clips. The training and validation sets of Kitchen20 dataset contain 480 and 160 clips, corresponding to the first three folds and the fourth fold, respectively. TriA$_{\mathrm{Kitchen20}}$ is divided into training and validation sets with a 4:1 ratio.

The Nonspeech7k dataset provides 6289 training and 725 test clips. We further split the training clips into training and validation sets with a 9:1 ratio. TriA$_{\mathrm{Nonspeech7k}}$ is also split into training and validation sets at a 9:1 ratio. The test set for the Nonspeech7k task is the Nonspeech7k test set.

\subsection{Results}

\begin{table}[t]
    \centering
    \caption{Experimental results on three AC tasks. A + B denotes sequential fine-tuning, where the model is first fine-tuned on A and then further fine-tuned on B.}
    \label{tab:4}
    \vspace{-0.2cm}
    \resizebox{\columnwidth}{!}{%
    \begin{tabular}{cccc}
        \toprule
        \textbf{Task} & \textbf{Dataset} & \textbf{Acc} & {\textbf{F1}}\\ 
        \midrule
        & DESED$_{\mathrm{real}}$ & 0.7837 & 0.7943 \\
        DESED$_{\mathrm{AC}}$ & TriA$_{\mathrm{DESED}}$ & 0.8255 & 0.7810 \\
        & TriA$_{\mathrm{DESED}}$ + DESED$_{\mathrm{real}}$ & \textbf{0.8258} & \textbf{0.8256} \\
        \midrule
        & Kitchen20 & 0.9250 & 0.9272 \\
        Kitchen20 & TriA$_{\mathrm{Kitchen20}}$ & 0.9375 & 0.9355 \\
        & TriA$_{\mathrm{Kitchen20}}$ + Kitchen20 & \textbf{0.9813} & \textbf{0.9812} \\
        \midrule
        & Nonspeech7k & 0.9448 & 0.9437 \\
        Nonspeech7k & TriA$_{\mathrm{Nonspeech7k}}$ & 0.8938 & 0.8734 \\
        & TriA$_{\mathrm{Nonspeech7k}}$ + Nonspeech7k & \textbf{0.9490} & \textbf{0.9464} \\
        \bottomrule
    \end{tabular}%
    }
\end{table}

Table~\ref{tab:4} reports the experimental results on three AC tasks. The model fine-tuned on TriA$_{\mathrm{DESED}}$ achieves higher accuracy than that fine-tuned on DESED$_{\mathrm{real}}$, but obtaining a lower Macro-F1. It suggests that TriA$_{\mathrm{DESED}}$ provides better overall data quality but exhibits slightly lower consistency across different classes. When applying sequential fine-tuning (TriA$_{\mathrm{DESED}}$ → DESED$_{\mathrm{real}}$), the model significantly outperforms the one fine-tuned solely on DESED$_{\mathrm{real}}$, achieving relative gains of 5.37\% in accuracy and 3.94\% in F1. It indicates that TriA$_{\mathrm{DESED}}$ helps the pre-trained model acquire task-relevant knowledge, and then DESED$_{\mathrm{real}}$ further improves the performance of the model.

TriA$_{\mathrm{Kitchen20}}$ outperforms Kitchen20 in both overall data quality and consistency across different audio classes. Sequential fine-tuning (TriA$_{\mathrm{Kitchen20}}$ → Kitchen20) achieves relative improvements of 6.09\% in accuracy and 5.82\% in Macro-F1 compared to fine-tuning on Kitchen20 alone, demonstrating that TriA$_{\mathrm{Kitchen20}}$ can help the model to learn the Kitchen20 task.

TriA$_{\mathrm{Nonspeech7k}}$ performs worse than Nonspeech7k in both accuracy and Macro-F1. However, sequential fine-tuning (TriA$_{\mathrm{Nonspeech7k}}$ → Nonspeech7k) still leads to performance gains over fine-tuning only on Nonspeech7k, with relative gains of 0.44\% in accuracy and 0.29\% in F1. It further demonstrates that TriA$_{\mathrm{GK}}$ can help the model learn the specific AC tasks.

Overall, fine-tuning on pipeline-processed data achieves performance comparable to, and in some cases better than, that obtained using manually annotated data. Compared to fine-tuning on manually annotated data alone, combining TriA$_{\mathrm{GK}}$ subsets with manual data through sequential fine-tuning achieves average relative improvements of 3.97\% in accuracy and 3.35\% in Macro-F1. It validates the effectiveness of TriA$_{\mathrm{GK}}$ for domestic AC tasks and further confirms the feasibility and practical value of the TriA Pipeline.

\section{Conclusion}

In this paper, we propose the TriA Pipeline, a large-scale automatic audio annotation pipeline. It efficiently converts audio collected from various streaming platforms into high-quality training data with event annotations. Based on the TriA Pipeline, the TriA dataset is constructed, which contains over 2130 hours of audio data covering 431 audio classes. Through comparative experiments, we verify the effectiveness of the prior-knowledge-guided subset, TriA$_{\mathrm{GK}}$, on domestic AC tasks, and further confirm that the TriA Pipeline is effective. The TriA Pipeline code and dataset is now released\footnote{\href{https://github.com/huanxian/TriA}{https://github.com/huanxian/TriA}}.

\section{Acknowledgments}

This work was partly supported by the national natural science foundation of China (62371195, 62111530145), and the exchange project of the 10th Meeting of China-Croatia Science and Technology Cooperation Committee (10-34).

\section{Generative AI Use Disclosure}

We used GPT-5.2 to assist in polishing the manuscript.

\bibliographystyle{IEEEtran}
\bibliography{mybib}

\end{document}